\documentstyle[leqno]{article}

\newcommand{\thenewsection}{\bf{\thesection}\rule{1em}{0em}}
\newenvironment{newsection}{\begin{flushleft}\rule{0mm}{3ex}\thenewsection
\bf}{\rule{0mm}{3ex}\noindent\end{flushleft}}

\newcommand{\mysection}[1]{\refstepcounter{section}
\begin{newsection}{#1}\end{newsection}}

\newcommand{\thenewsubsection}{\bf{\thesubsection}\rule{1em}{0em}}
\newenvironment{newsubsection}{\begin{flushleft}\rule{0mm}{3ex}\thenewsubsection
\bf}{\rule{0mm}{3ex}\noindent\end{flushleft}}

\newcommand{\mysubsection}[1]{\refstepcounter{subsection}
\begin{newsubsection}{#1}\end{newsubsection}}

\newcommand{\proof}{{\em Proof.\ }}

\newcommand{\qed}{\hspace*{\fill}\rule{2mm}{2mm}\vspace{5mm}}

\newcommand{\norm}[1]{\mbox{$\left\|{#1}\right\|\,$}}

\newcommand{\ra}{\rightarrow}

\newcommand{\arr}[1]{\stackrel{#1}{\longrightarrow}}

\newcommand{\larray}{\left(\begin{array}{cc}}
\newcommand{\rarray}{\end{array}\right)}

\newcommand{\beq}{\begin{equation}}
\newcommand{\eeq}{\end{equation}}
\newcommand{\barr}{\begin{array}}
\newcommand{\earr}{\end{array}}
\newcommand{\beqar}{\begin{eqnarray}}
\newcommand{\eeqar}{\end{eqnarray}}

\newtheorem{proposition}{Proposition}

\newtheorem{leftnumbers}{\rule{-0.35em}{0em}}[section]
\newenvironment{themit}[1]{\begin{leftnumbers}{\sc #1.}}{\end{leftnumbers}}

\newcommand{\btheorem}{\begin{themit}{Theorem}\ }
\newcommand{\etheorem}{\end{themit}}
\newcommand{\bproposition}{\begin{themit}{Proposition}\ }
\newcommand{\eproposition}{\end{themit}}
\newcommand{\blemma}{\begin{themit}{Lemma}\ }
\newcommand{\elemma}{\end{themit}}
\newcommand{\bconjecture}{\begin{themit}{Conjecture}\ }
\newcommand{\econjecture}{\end{themit}}
\newcommand{\bproblem}{\begin{themit}{Problem}\ }
\newcommand{\eproblem}{\end{themit}}
\newcommand{\bcorollary}{\begin{themit}{Corollary}\ }
\newcommand{\ecorollary}{\end{themit}}

\newenvironment{remdefexam}[1]{\begin{leftnumbers}{\sc #1.}
 \rm}{\end{leftnumbers}}

\newcommand{\bremark}{\begin{remdefexam}{Remark}\ }
\newcommand{\eremark}{\end{remdefexam}}
\newcommand{\bdefinition}{\begin{remdefexam}{Definition}}
\newcommand{\edefinition}{\end{remdefexam}}
\newcommand{\bexample}{\begin{remdefexam}{Example}\ }
\newcommand{\eexample}{\end{remdefexam}}
\newcommand{\bexercise}{\begin{remdefexam}{Exercise}\ }
\newcommand{\eexercise}{\end{remdefexam}}

\renewcommand\theequation{\thesection.\arabic{leftnumbers}}
\newcommand{\beqn}{\refstepcounter{leftnumbers}
\begin{equation}}
\newcommand{\eeqn}{\end{equation}}

\newtheorem{remit}[proposition]{Remark}

\newtheorem{defineit}[proposition]{Definition}

\newtheorem{examit}[proposition]{Example}

 \newcounter{subeqn}[equation]
\renewcommand{\thesubeqn}{\theequation.\alph{subeqn}}

\newcommand{\cala}{\mbox{$\cal A$}}
\newcommand{\calb}{\mbox{$\cal B$}}

\newcommand{\cale}{\mbox{$\cal E$}}

\newcommand{\calg}{\mbox{$\cal G$}}
\newcommand{\calh}{\mbox{$\cal H$}}

\newcommand{\call}{\mbox{$\cal L$}}

\newcommand{\calp}{\mbox{$\cal P$}}

\newcommand{\cals}{\mbox{$\cal S$}}

\newcommand{\calu}{\mbox{$\cal U$}}

\newcommand{\bc}{{\bf C}}

\newcommand{\bh}{{\bf H}}

\newcommand{\bz}{{\bf Z}}

\newcommand{\mobius}{M\"{o}bius\ }
\newcommand{\fm}{\mbox{\rm FM}}
\newcommand{\union}{\bigcup}
\newcommand{\Tr}{\mbox{\rm Tr}}
\newcommand{\half}{\frac{1}{2}}

\title{M\"{o}bius Transformations in Noncommutative Conformal Geometry}
\author{P.~J.~M.~Bongaarts\thanks{Supported in part by the 
Visitor Fund of the Department of Mathematics, University of Exeter.}\\
\footnotesize \em  Lorentz Institute for Theoretical Physics\\
\footnotesize \em Leiden University \\
\footnotesize \em P.O. Box 9506\\
\footnotesize \em 2300 RA Leiden, The Netherlands\\
{\footnotesize\tt bongaart@lorentz.leidenuniv.nl}
 \and 
J.~Brodzki
\thanks{Supported in part by an LMS Scheme 4 grant and a grant from the Exeter 
University Research Fund.}\\
\footnotesize \em Department of Mathematics\\
\footnotesize \em University of Exeter\\
\footnotesize\em North Park Road \\
\footnotesize \em Exeter EX4 5HE, UK\\
{\footnotesize\tt brodzki@maths.ex.ac.uk}}
\date{}
\begin{document}

\maketitle
\begin{abstract}
We study the projective linear group $PGL_2(A)$  associated with an arbitrary 
algebra $A$, and  its subgroups from the point 
of view of their action on the space of involutions in $A$. This action
formally resembles
 \mobius transformations known from complex geometry.
By specifying $A$ to be an algebra of bounded operators 
in a Hilbert space $\bh$, we rediscover  the \mobius group $\mu_{ev}(M)$ 
defined by Connes and study its action on the space of Fredholm modules
over the algebra $A$. There is an induced action on the $K$-homology
of $A$, which turns out to be trivial. Moreover, 
this action leads naturally to a simpler object, 
the polarized module underlying a given Fredholm module, 
and we discuss this relation in detail. Any polarized module can be lifted 
to a Fredholm module, and the set of  different lifts forms 
a category, whose morphisms are given by 
generalized \mobius tranformations.  We present an example
of a polarized module canonically associated with the differentiable 
structure of a smooth manifold $V$. Using our lifting procedure
we obtain a class of Fredholm modules characterizing
the conformal structures on $V$. Fredholm modules obtained in this 
way are a special case of those constructed by Connes, Sullivan and Teleman.
\end{abstract}

\vfill \eject

\begin{center}
\bf Introduction
\end{center}
Much of the pioneering work of Connes in noncommutative geometry rests on the 
notion of Fredholm module, which arises as an abstract formulation of
the calculus of elliptic operators on a differentiable manifold. It has been 
demonstrated in numerous examples that Fredholm modules contain much of 
the geometric information associated with the underlying manifold. 
A key result in this direction is the theorem of Connes, Sullivan and 
Teleman \cite{CST1}\cite{CST2} which associates an even Fredholm module with 
each oriented even dimensional conformal manifold. Conversely, given such a 
Fredholm module, one can reconstruct the original conformal structure 
\cite[p.332]{Connes:Book}.
It thus transpires that Fredholm modules are very close to being 
noncommutative analogues of conformal structures on differentiable 
manifolds. This fact indicates that 
they play a more basic role than spectral triples that capture the essence 
of the Riemannian geometry of smooth manifolds. As is well known, 
conformal structure is fundamental in general relativity, where it carries
 information about causality. Moreover, many interesting non-compact conformal
 manifolds can be conformally compactified (unlike the Riemannian case)
 which is a further indication that Fredholm modules will be useful in 
 this context. They provide also a natural framework for the study of pseudo-Riemannian
 structures in noncommutative geometry. 

The point of departure for our work is a result of Connes, 
which introduces a group of transformations of
Fredholm modules. These transformations are formally analogous to 
\mobius transformations known from complex geometry. 
In order to provide geometric background for them  we define, for any algebra $A$,  
the projective line $P_1(A)$ and the projective group $PGL_2(A)$.
This group contains several interesting subgroups that act in various
ways on the set of involutions in $A$. One of the most interesting
cases in which this construction may be considered is 
when $A$ is an algebra of bounded operators  on 
a Hilbert space and $M$ is its commutant. 
 In this case it is possible to give a complete 
descriptions of subgroups of $PGL_2(A)$ that act on the space of 
self-adjoint involutions. Among those subgroups we discover the 
\mobius 
group $\mu_{ev}(M)$ defined by Connes.
We study the structure of this group in some detail. In particular, 
we employ the Cayley transform to rederive the general formula for the 
polar decomposition of \mobius transformations. From this it 
follows easily that the \mobius group retracts  onto the group of unitary elements in $M$. 
Since the same result is true for any $C^*$-algebra $A$, we can combine this
statement with our further study  of the action of the \mobius group on the 
space of Fredholm modules to prove that \mobius transformations act trivially 
on the $K$-homology of the algebra $A$. 

After this basic set-up is ready, we turn to the question of the action
of the \mobius group on the space  of Fredholm modules.  We first 
study in detail the structure of the space of unitarily equivalent Fredholm 
modules, proving a \lq localization theorem'. We then  discuss the action of the
 \mobius group. 
An interesting feature of the action of this  group on the space
of Fredholm modules is that it gives rise to a simpler object, 
the polarized module. Roughly speaking, Fredholm modules contain
information about confomal structure of a space, whereas polarized
modules encode its \lq differential\rq\ structure. We study in detail 
the relation between these objects and show in particular (building on the 
results of \cite{BCE}) that 
any Fredholm module has a canonical underlying polarized module. Conversely, 
any polarized  module may be lifted to a Fredholm module. This lift is not
unique, in general, and in fact we prove that all possible Fredholm 
module lifts of a given polarized module may be conveniently grouped 
together in a category, whose morphisms are generalized \mobius transformations.
The most important example of a polarized module is one that can be 
canonically associated with the differentiable structure of a smooth 
manifold $V$. Any lift of this canonical polarized module to a Fredholm 
module depends on the choice of a conformal structure on $V$.
Any Frehdolm module obtained this way turns out to be a special case
of the Connes-Sullivan-Teleman  Fredholm module associated
with a general conformal manifold of even dimension.

Another insight gained from the study of geometric examples is 
that Fredholm modules are normally constructed using algebras of complex-valued
functions acting in  a complex Hilbert space. Thus if one is interested 
in real manifolds and algebras real-valued functions, it seems that part of the 
data is lost in the process. This suggests that there should 
be a property of Fredholm modules that would describe 
this real structure. Hence we introduce
the notion of  real Fredholm modules and real polarized modules and 
show that the action of the \mobius group preserves
this notion. The canonical polarized modules and its lifts to a Fredholm 
module are real in this sense. 

\mysection{The group of \mobius transformations}
\mysubsection{The projective line $P_1(A)$  and the projective group $PGL_2(A)$}
In this section,  we  set up  the algebraic background for  
discussion of projective lines and projective groups associated with an
algebra $A$. The following section will bring topological considerations, 
where we shall assume that the algebra $A$ is a Banach or a $C^*$-algebra. 

Let $A$ be a complex unital algebra.  
We denote by  $GL_1(A)$ the group of invertible elements in $A$.
Let us introduce the following equivalence relation 
on the space  $A \times A - \{(0,0)\}$. We say that 
$(a_1, b_1) \sim (a_2, b_2)$, 
for $a_i, b_i \in A$ iff there exists $\lambda \in GL_1(A)$ such that 

$$
  a_1 = a_2 {\lambda}, \qquad  b_1 =
b_2{\lambda}
$$

\bdefinition 
The quotient space $(A\times A- \{(0,0)\})/\sim$,  
denoted $P_1(A)$, will be  called the {\em projective line} 
over $A$ .
\edefinition
\bremark
If the algebra $A$ is not commutative, then there are  two ways
to define the projective line, corresponding to the left and right actions of 
the group $GL_1(A)$ on $A\times A$. 
\eremark
\bdefinition 
The {\em finite part},
or set of {\em finite points} of $P_1(A)$,
consists of the equivalence classes of pairs $(a,b)$ in which $b$ is invertible.
This set will be  denoted ${P_1(A)}^f$.
\edefinition 
\bremark
 A representative  $(a,b)$ of  an equivalence class
in ${P_1(A)}^f$ is clearly equivalent to the pair $(ab^{-1}, 1)$.
This implies that there is a $1-1$ correspondence
between elements of ${P_1(A)}^f$ and $A$,
given by $[(a,b)] \mapsto z = ab^{-1}$.
\eremark
\bremark 
In the case where $A$ is the field of complex numbers,
the projective line is obtained
from the set of \lq finite' complex numbers
by adding a single \lq point at infinity'.
In the general case
one may have many elements in the algebra $A$ that are non-zero,
but nevertheless not invertible,
so the projective line is obtained there by adding not just a single point,
but a {\em set} of \lq points at infinity'.
\eremark

Let $M_2(A)$ be the algebra of $2 \times 2$ matrices with entries in $A$,
and let $GL_2(A)$ be the group of invertible elements of $M_2(A)$. Both
$M_2(A)$ and $GL_2(A)$ act on the left  on the space $A\times A$ by 
the usual matrix multiplication. Since this left action obviously commutes
with the right diagonal action of the group $GL_1(A)$ introduced earlier, 
the action of  $GL_2(A)$ descends to  the projective line $P_1(A)$. 

The induced action of $GL_2(A)$ on $P_1(A)$ is not effective, as is not difficult to see. 
Let $Z(A)$ be the center of the algebra $A$, 
i.e, $Z(A) = \{ a \in A \mid  ab =ba, \forall b \in A \}$.
It is clear that elements of $GL_2(A)$ of the form
\beqn\label{effective-action}
\left(
\begin{array}{cc}
a & 0\\
0 & a
\end{array}
\right)
\eeqn
with
$a \in Z(A) \cap GL_1(A)$ leave any point of the projective line $P_1(A)$ 
invariant. 
\blemma
Let $N$ be the subgroup of $GL_2(A)$ with the property that 
$g(x) = x$ for any $g\in N$ and $x\in P_1(A)$. Then 
$$
N = \left\{ \left(
\begin{array}{cc}
a & 0\\
0 & a
\end{array}
\right) \mid a\in  Z(A) \cap GL_1(A)\right\}.
$$
\elemma
\proof  Let $T \in GL_2(A)$ with matrix entries $a,b,c,d$,
be  such that for every $(x_1,x_2) \in A \times A- \{(0,0)\}$,
there exist a ${\lambda} \in GL_1(A)$
such that $T(x_1,x_2) = (x_1{\lambda}, x_2{\lambda})$.
Taking for $(x_1,x_2)$ successively
$(0,1)$, $(1,0)$ and $(1,1)$
one obtains $b=0$, $c=0$ and $d=a$.
Next,  take a pair $(1, x)$, where   $x \in A$ is arbitrary.
There exists a ${\lambda}_x \in GL_1(A)$
such that $T(1,x) = ({\lambda}_x, x{\lambda}_x)$,
which gives immediately ${\lambda}_x = a$
and then $ax=xa$. So $ax =xa$ for arbitrary $x \in A$.
Finally, a diagonal matrix $T \in GL_2(A)$ whose diagonal entries both 
equal $a$
is invertible iff $a$ is invertible. \qed

Since it is clear that $N$ is a normal subgroup of $GL_2(A)$, we can 
introduce the following definition. 
\bdefinition
The projective linear group $PGL_2(A)$ is by definition the quotient
group 
$$
PGL_2(A) = GL_2(A)/N
$$
\edefinition
It follows from the Lemma that the projective group $PGL_2(A)$ 
acts effectively on the projective line $P_1(A)$. 

We shall now study various subgroups of $PGL_2(A)$. 
Our first order of business will be to identify elements from $PGL_2(A)$ which 
restrict,  in some sense, to transformations of the subset $P_1(A)^f $ 
of finite points of the projective line.

Let then  $T \in PGL_2(A)$ be represented by a matrix
$T= \larray a & b \\ c & d
\rarray$ with entries in $A$. 
Since $T$ acts on $P_1(A)^f$ by 
$$
\larray a & b \\ c & d \rarray \left( \begin{array}{c}
z \\ 1\end{array}\right) = \left( \begin{array}{c}
az + b \\ cz +d \end{array}\right) 
$$
the vector on the right is an element of $P_1(A)^f$ if and 
only if 
$$
\left( \begin{array}{c}
az + b \\ cz +d \end{array}\right)\sim \left( \begin{array}{c}
(az + b)(cz +d)^{-1}\\ 1 \end{array}\right)
$$
This happens when 
$cz +d$ is invertible for any $z\in P_1(A)^f$. As this assumption 
is quite restrictive, we shall first investigate possible 
action of \mobius transformations on the subset of involutions 
in $P_1(A)^f$, i.e., elements $f$ such that $f^2 = 1$. We begin 
with the simplest possible case. 

\blemma\label{3.9}
 Let $T \in PGL_2(A)$ be as above. Then the action of $T$ 
 on the two trivial involutions in $A$,
$f_+=+1$ and $f_-= -1$ is well defined
if and only if
$c+d$ and $c-d$ are invertible. Furthermore, $Tf_{\pm} = f_{\pm}$ 
if and only if $c=b$ and $d=a$.
\elemma

\proof The first part of the  lemma follows immediately
from the application of the formula
$$
z \mapsto (az+b)(cz + d)^{-1}
$$
for the action of $T$ on  $P^f_1(A)$.
Note that the independence of the choice of representative
in the equivalence class of  $T$ is obvious. 

For the second part,  applying the above formula one gets
$a+b=c+d$ and $-a+b=c-d$.
Addition and subtraction then gives $c=b$ and $d=a$.
Independence of choice of representative is again obvious.

\bremark As projective transformations are 
not linear in general,  the  conditions for the action of $T$ on 
$f_{\pm}$ are independent. 
\eremark

Let us denote by $\calg$ the set of matrices $\larray a & b \\
b & a \rarray\in GL_2(A) $ in which $a+b$ and $a-b$ are invertible. 

\bproposition \label{special-Mobius}
$\cal{G}$ is a subgroup of $GL_2(A)$, and its image $G$ under the 
canonical surjection $GL_2(A) \ra PGL_2(A)$ 
is a  subgroup of the projective linear group $PGL_2(A)$.
\eproposition
\proof The identity matrix  clearly is an element of $\cal{G}$,
and a simple calculation shows
that a product of two matrices from $\cal{G}$
is again a matrix in $\cal{G}$.
An element $T \in {\cal G}$ is by definition
an element of $GL_2(A)$ and therefore has
an inverse $T^{-1}$ in $GL_2(A)$.
It is clear that $T^{-1}$ is defined on the two trivial involutions $f_{\pm}$
and leaves them invariant, so $T^{-1}$ is also in $\cal{G}$,
because of Lemma \ref{3.9}.
Moreover, the kernel of the canonical  homomorphism of $GL_2(A)$
onto $PGL_2(A)$ is a normal subgroup of $\cal{G}$.
Therefore the quotient group of $\cal{G}$
can be identified with a subgroup $G$ of the  group $PGL_2(A)$.
\qed

\bproposition\label{first-isomorphism}
The group $\cal{G}$ is isomorphic to the direct product group
$GL_1(A)\times GL_1(A)\subset GL_2(A)$. The subgroup $G$ of the 
projective linear group $PGL_2(A)$ is isomorphic to 
$(GL_1(A)\times GL_1(A))/N$. 
 \eproposition
\proof Define $x$ and $y$ in $A$ as
$$
\begin{array}{rcl}
x & = & a+b\\
y & = & a-b
\end{array}
$$
with the inverse relations
$$
\begin{array}{rcl}
a & = & \half (x+y)\\
\rule{0mm}{5mm}
b & = & \half (x-y)
\end{array}
$$
The invertibility of $a+b$ and $a-b$
is  equivalent to the invertibility of $x$ and $y$.
Matrix multiplication expressed in these coordinates becomes
 $(x_1,y_1)(x_2, y_2) = (x_1x_2, y_1y_2)$, where $x_i, y_i \in GL_1(A)$, 
which means that the group $\cal{G}$ is isomorphic
to the direct product group ${GL_1(A) \times GL_1(A)}$.
Although not necessary here, we mention that using 
this parametrisation it is also easy to find the 
inverse of a matrix in $\cal{G}$. Since the 
inverse of $(x,y)$ is $(x^{-1}, y^{-1})$ we 
get that 
\beqn \label{inverse}
\rule{3mm}{0mm}
\larray a & b \\ b & a \rarray ^{-1} 
= \frac{1}{2}\larray  (a+b)^{-1} + (a-b)^{-1} &  (a+b)^{-1} - 
(a-b)^{-1} \\
(a+b)^{-1} - 
(a-b)^{-1} & (a+b)^{-1} + (a-b)^{-1}
\rarray
\eeqn
To prove the statement about the subgroup $G$ of $PGL_2(A)$, 
it is sufficient to note that $N$ is a normal subgroup of $GL_1(A)\times
GL_1(A)$. 
\qed

We finish this discussion by recording one more simple fact. 
\bproposition\label{image-is-involution}
Let $T\in G$. Assume that $T$ is represented by a matrix 
$T = \larray a & b \\ b & a \rarray$. 
If, for some  involution $f\in A$, $bf+a$ is invertible, 
then the image of $f$ under  $T$ is an involution. 
\eproposition
\proof 
 We use the formula
$f \mapsto (af+b)(bf + a)^{-1}$, which is independent of the choice 
of representative of $T$ in $GL_2(A)$.
Because  $f^{-1} = f$,  this can be written as
$f \mapsto (af+b)((b +af^{-1})f)^{-1} = (af+b)f(af+b)^{-1}$,
which immediately gives $[(af+b)(bf+a)^{-1}]^2 = 1$. \qed

\mbox{}

Let us now be a bit more daring and consider projective transformations
$T$ that are defined on {\em all} involutions $f\in A$. 
We start with the subgroup $\cal{G}_C$ of $GL_2(A)$
consisting of matrices   $\larray a & b \\ b & a \rarray
$ such that $bf + a$ is invertible for any involution $f$ in 
$A$. We shall denote by $G_C$ the image of ${\calg}_C $ in $PGL_2(A)$. 

\bproposition
$\cal{G}_C$ is a subgroup of $GL_2(A)$, and 
$G_C$ is a subgroup of $PGL_2(A)$ acting effectively on the set of all 
involutions in $A$. 
\eproposition
\proof Let $T_i\in \cal{G}_C$  be given by matrices 
$\larray a_i & b_i \\ b_i & a_i \rarray $, $i= 1,2$. We need to prove 
that the product $T_2T_1$ is an element of $\cal{G}_C$, 
i.e., that $bf +a$ is invertible for all involutions $f$, where 
$$
a = a_2a_1 + b_2 b_1, \qquad b = a_2 b_1 + b_2 a_1.
$$
Let $f$ be an involution in $A$ and let $f_1 = (a_1f + b_1)(b_1f + a_1)^{-1}$. 
Then it is  easy  to check that 
$$
bf + a = (b_2f_1 + a_2)(b_1f + a_1)
$$
hence it is invertible as the product of two invertible elements. 
Finally, it is clear that $N$ is a normal subgroup of $\cal{G}_C$, and 
so the second statement follows.  
\qed

To summarize, we have identified subgroups of $GL_2(A)$ and 
$PGL_2(A)$ that fit into the following commutative diagram  

\beqn\label{commutative-diagram}
\begin{array}{ccccccc}
N & \arr{} & \cal{G}_C & \arr{} &  \cal{G} & \arr{} & GL_2(A) \\
  &        & \downarrow &        &\downarrow &      & \downarrow \\
  &        &   G_C       & \arr{} &   G      & \arr{}& PGL_2(A)
\end{array}
\eeqn
in which the horizontal arrows are inclusions and vertical arrows 
are surjections. 
The group $G$ acts effectively on the projective line $P_1(A)$, leaving the 
two trivial involutions $f_{\pm}$ fixed, whereas the subgroup $G_C$  
acts on the set of all involutions in $A$. 

\mbox{}

Let us now assume that the algebra $A$ is $\bz_2$-graded. This grading
is inherited by the matrix algebra $M_2(A)$, and in particular even 
matrices in $M_2(A)$ are those
whose diagonal entries are even and off-diagonal entries are
odd elements of $A$. 

Using this additional structure we can single out a new subgroup 
of  $\calg$, namely $\calg^{ev} = \calg \cap 
M_2^{ev}(A)$. This group consists of all {\em even} matrices 
$\larray a & b \\ b & a \rarray$ in which $a+b$ and $a-b$ 
are invertible. These matrices preserve the trivial involutions
$\pm 1$. Similarly, we have the subgroup $\calg_C^{ev} = \calg_C 
\cap M_2^{ev}(A)$. It is straightforward to check that
elements of the group $\calg_C^{ev}$ map odd involutions in 
$A$ to odd involutions. 

As before, the action of the groups $\calg^{ev}$ and $\calg_C^{ev}$
is not effective, as the elements of $N^{ev}= Z(A) \cap GL_1(A)\cap 
M_2^{ev}(A)$ act as identity transformations on involutions. 
The canonical surjection $GL_2(A) \ra PGL_2(A)$ restricts to 
$GL_2(A)^{ev}$ and
the kernel of this restriction is $N^{ev}$ thus 
giving a new group $PGL_2^{ev}(A)$. Note that we do not
mean to imply here that $PGL_2(A) $ is a graded group.  
$N^{ev}$  is a normal 
subgroup of both $\calg^{ev}$ and $\calg_C^{ev}$. Thus we get two 
subgroups $G_C^{ev}$ and $G^{ev}$ of $PGL_2^{ev}(A)$.
$G_C^{ev}$ acts effectively on the set of involutions, 
sending odd involutions to odd involutions.  These new
groups form a commutative diagram identical to \ref{commutative-diagram}.

\mysubsection{The projective group and $*$-algebras}

From this point onwards we shall assume that $A$ is a $*$-algebra (which 
will become a Banach or a $C^*$-algebra later on),
so that one can speak of selfadjoint and unitary elements of $A$. 
In the presence of this
additional structure, our initial observations concerning the groups
$\cal{G}$, $\cal{G}_C$, $G$ and $G_C$ can be sharpened. In particular, 
we set out to investigate under what additional conditions the groups 
$\cal{G}_C$ and $G_C$ will act on the set of self-adjoint involutions 
in $A$. 

Let us begin with the subgroup $\cal{G}_C$ of matrices
$\larray a & b\\ b & a \rarray 
$  such that $bf +a$ is invertible for all involutions 
$f$ in $A$.
We are looking for conditions on $a,b$ that would ensure that 
the involution $f' = (af + b)( bf+a)^{-1}$ is self-adjoint, for any self-adjoint
involution $f$. 

\blemma\label{commutation}
The image $f'= (af + b)(bf + a)^{-1}$ of a  self-adjoint involution 
$f$ is  self-adjoint if and only if the following identities hold. 
\beqn\label{commutation-evidence}
\begin{array}{rcl}
[a^*a - b^*b , f ] & = & 0 \\ \mbox{}
[a^*b - b^* a , f ] & = & 0 \\ \mbox{}
[aa^* - bb^* , f ] & = & 0 \\  \mbox{}
[ ab^* - ba^*, f ] & = & 0 
\end{array} 
\eeqn 
\elemma
\proof We have already seen that $f'$ is an involution. It will be self-adjoint
iff
$$
((af + b)( bf + a)^{-1})^* = (af + b )(bf + a)^{-1}
$$
which is equivalent to 
$$
(fa^*+ b^*)(bf + a ) = ( a^* + fb^*)(af+b). 
$$
We rewrite the last formula as 
$$
f(a^*b - b^*a) f - [f, a^*a - b^*b] - ( a^*b - b^*a) = 0
$$
and note that, since it is expected to hold for all 
self-adjoint involutions $f$ it must, in particular, hold 
for $-f$. Proceeding in the same way with $-f$ in place of $f$ 
produces a similar condition, and the two together yield
$$
\begin{array}{rcl}
[a^*a - b^*b , f ] & = & 0 \\ \mbox{}
[a^*b - b^* a , f ] & = & 0
\end{array}
$$
To derive the latter two, we  use   variables $x = a+b$ and
$y= a-b$  to rewrite the last two formulae in the following 
form
$$
\begin{array}{rcl}
[x^*y + y^*x, f] & = & 0\\ \mbox{}
[x^*y - y^*x , f] & = & 0 
\end{array}
$$
which leads to a single condition $[x^*y, f]=0$. The inverse transformation
to $f\mapsto f'$, which depends on  $x^{-1}$ and $y^{-1}$, leads to 
a similar condition $[(x^*)^{-1} y^{-1}, f] = 0$ which is equivalent to 
$$
[xy^*, f]=0. 
$$
This in turn gives
$$
\begin{array}{rcl}
[aa^* - bb^*, f] & = & 0  \\ \mbox{}
[ab^* - ba^*, f] & = & 0. 
\end{array}
$$
This finishes the proof of the lemma. \qed
\bremark
It is clear that the identities \ref{commutation-evidence} still 
hold if $a$ and $b$ are replaced by $au=ua$, $bu=ub$, with $u\in 
Z(A)\cap GL_1(A)$, and so they descend to the group $G_C\subset 
PGL_2(A)$. 
\eremark

In the case when $A$ is the $C^*$-algebra  $\call$ 
of bounded operators in a Hilbert space $\bh$ we can say a bit more. 
Since to any self-adjoint involution $f$ corresponds a self-adjoint projection 
$e = (1+f)/2$, the formulae \ref{commutation-evidence} imply that the same
identities should hold for an arbitrary self-adjoint projection. Thus the 
operators on the left in \ref{commutation-evidence} have to be multiples 
of the identity operator on $\bh$. We thus have that
\beqn\label{Schur}
\begin{array}{rcl}
a^*a - b^* b & = & \lambda _1\\
a^*b - b^* a & = & i\lambda_2 \\
aa^* - bb^* & = & \lambda_3\\
ab^* - ba^* & = & i\lambda_4
\end{array}
\eeqn
where $\lambda_i$ are real numbers.
Rewriting these formulae in terms of $x$ and $y$ we find 
$$
\begin{array}{rcl}
x^*y & = & \lambda_1 - i\lambda_2\\
xy^* & = & \lambda_3 - i\lambda_4, 
\end{array}
$$
or, if we use that $x$ and $y$ are invertible, 
$$
\begin{array}{rcl}
x &=& (\lambda_3 + i\lambda_4)(y^*)^{-1}\\
y & = & (\lambda_1 - i\lambda_2)^{-1}(x^*)^{-1}
\end{array}
$$
This gives that $\lambda_1 = \lambda_3$ and $\lambda_2 = - \lambda_4$ with 
$\lambda_1$ and $\lambda_2$ not both equal zero. 

Our calculations thus demonstrate that elements of the group ${\cal{G}}_C$ 
may be written 
in the $x,y$ variables
as $(x, \alpha(x^*)^{-1})$, where $x $ is an arbitrary invertible 
element of $A$ and $\alpha = \lambda_1 -
i\lambda_2$ is an arbitrary complex number. Any such pair can be written as
$$
(x, \alpha (x^*)^{-1}) = (1, \alpha )(x, (x^*)^{-1})
= (x, (x^*)^{-1}) (1,\alpha). 
$$
This leads to the following statement. 
\btheorem
Let $A$ be the algebra of bounded operators on a Hilbert space $\bh$. 
Then the group $\cal{G}_C$ 
is the direct product ${\cal{G}}_C = GL_1(A)\times \bc^\times$
of the group of invertible elements in $A$ and the multiplicative group of
complex numbers. The image $G_C$  of this group in $PGL_2(A)$ is 
the group $PGL_1(A) = GL_1(A)/\bc^\times$. 
\etheorem
\proof The first part of the theorem has already been proved. The 
statement about the group $G_C$ follows from the fact that,  
if $A$ is the algebra of bounded operators on a Hilbert 
space, then by Schur's lemma its centre is just the field of 
complex numbers, so that $Z(A) \cap GL_1(A) = \bc^\times$, in this case. 
\qed

Keeping our assumption that $A$  is the algebra 
of bounded linear operators on a Hilbert space we note that the 
group $\cal{G}_C$ contains an interesting subgroup, which is isomorphic to 
$GL_1(A)$ and which  consists of elements $(x, (x^*)^{-1})$, $x\in A$. 
This  corresponds
to the choice of $\lambda_1 = \lambda_3 = 1$ and $\lambda_2 =- \lambda_4 =0$. 
This is precisely  the group $\mu(A)$ 
defined by Connes \cite[p.~335]{Connes:Book}, which consists of 
matrices $ \larray a & b \\ b & a \rarray \in GL_2(A)$ such that 
\beqn\label{condition}
\rule{3mm}{0mm}
\larray a & b \\ b & a \rarray \larray a^* & -b^* \\ -b^* & a^*
\rarray = 
\larray a^* & -b^* \\ -b^* & a^*
\rarray
\larray a & b \\ b & a \rarray
 = \larray 1 & 0 \\ 0 & 1 \rarray 
\eeqn
We note that this condition is equivalent to Connes' identities
\beqn\label{Connes-group}
\begin{array}{rcl}
a^*a - b^* b & = & 1\\
a^*b - b^* a & = & 0 \\
aa^* - bb^* & = & 1\\
ab^* - ba^* & = & 0
\end{array}
\eeqn
and which follows directly from our computations. Moreover, as 
these identities imply that $(a+b)^{-1} = a^* - b^*$ and $(a-b)^{-1}= 
a^* + b^*$, it is clear that condition \ref{condition} is a special case
of the formula \ref{inverse}. Hence our results put Connes' construction
in an interesting geometric context. 
The group $\mu(A)$  will 
be our main object of study, with the algebra $A$ to be specified. 

Finally, let us assume that $A$ is a  $\bz_2$-graded algebra with 
grading given by conjugation by a fixed involution $\gamma$. Following
the procedure outlined here 
we uncover the group $\mu_{ev}(A)$ which is defined in the
same way as $\mu(A)$ with the additional requirement that $a$ be even 
and $b$ be odd in $A$. 
\bproposition
$\mu_{ev}(A)$ is isomorphic to the subgroup of $GL_1(A)$ consisting 
of elements $x$ such that $\gamma x^*\gamma = x^{-1}$. 
\eproposition
\proof \cite{BCE} Define a map $\mu_{ev}(A) \ra GL_1(A)$ 
by
$$
\larray a & b \\
b & a \rarray \mapsto x = a+b
$$
with $a$ even and $b$ odd. The identities \ref{Connes-group} give in this
case that 
$$
\gamma x^* \gamma = \gamma (a + b)^*\gamma = (a-b)^* = x^{-1}
$$
\qed

\mysubsection{The \mobius group over a $C^*$-algebra}
The general algebraic formalism developed in previous section 
becomes particularly useful when the algebra over which 
the \mobius group is defined is a $C^*$-algebra. We shall now look
into possible refinements of the algebraic theory under this 
additional assumptions. 
There will be two special cases that roughly determine the 
scope of interesting choices of an algebra. One is the 
essential commutant of the representation $\pi$, and the other
is the commutant of $\pi$. The first case is of interest from the 
point of view of $K$-homology, whereas the second is better suited
to geometric applications. One of the tools that we shall use
in further study of the \mobius group is the Cayley transform, 
and so we begin by recalling its basic properties. 

\bproposition\label{Cayley}
 Let ${\bh}$ be a (complex) Hilbert space.
There is a $1-1$ correspondence
between bounded invertible positive operators $Q$,
and bounded selfadjoint operators $m$ with
norm strictly smaller than one.
This correspondence is given by the formula
$$
Q = \frac{1 - {m}}{1+ {m}},
$$
with the inverse relation
$$
{m} = \frac{1-Q}{1+ Q}.
$$
The operator $Q$ is called the {\em Cayley transform} of $m$.
\eproposition

It is not difficult to see that the Cayley transform of $Q^{-1}$ is $-{m}$.
Note also that the 
fractional notation in the
above formulae is justified,
because
$$
(1-{m})(1+{m})=(1+{m})(1-{m})
$$
implies
$$
(1 + {m})^{-1}(1-{m}) = (1 - {m})(1+{m})^{-1}.
$$
Finally, we observe that $Q$ and ${m}$ commute.

\mbox{}

Let $A$ be a $C^*$-algebra of operators in a complex Hilbert 
space $\bh$. In this case the group $\mu(A)$, defined in the previous section, 
will be called {\em the \mobius group}. In the graded case 
we shall consider the even group $\mu_{ev}(A)$.
The isomorphism that identifies $\mu(A)$ with $GL_1(A)$, 
or a subgroup of $GL_1(A)$ in the graded case, allows one
to define unitary and positive elements on $\mu(A)$,  and to introduce polar decomposition of elements on the \mobius group. And so, an element 
$g= \larray a & b \\ b & a \rarray $ of the \mobius group is unitary
if and only if the corresponding element $x = a + b$ is unitary in 
$GL_1(A)$. Such a $g$ is said to be positive if and only if $x$ is
positive in $GL_1(A)$. 

\bproposition\label{unitary-and-positive}
An element  $g\in \mu(A)$ is unitary if and only if 
it has the form
$$
g=
\left(
\begin{array}{cc}
u & 0\\
0 & u
\end{array}
\right),
$$
with $u$ a unitary operator in $A$.

Such an element $g$  is positive if and only if
it has the form
$$
g =
\left(
\begin{array}{cc}
(1-{m}^2)^{-\frac{1}{2}} & -{m}(1-{m}^2)^{-\frac{1}{2}}\\
-{m}(1-{m}^2)^{-\frac{1}{2}} & (1-{m}^2)^{-\frac{1}{2}}
\end{array}
\right),
$$
with ${m}$ a bounded, selfadjoint operator in $A$
which has norm strictly smaller than one.
\eproposition
\proof The first part of the proposition is clear. For the second, 
let us denote the image of $g$ in $GL_1(A)$ by $x$. We 
assume that $x$ is positive. 
We use the Cayley transform \ref{Cayley} to define 
 a selfadjoint operator ${m}$,
which has norm strictly smaller than $1$,
by the formula
$$
 {m}= \frac{1-x^2}{1+x^2}.
$$
Continuous functional calculus on $A$ shows that $m$ is an element 
of $A$. Moreover, since $x$ is positive,
$$
x= \left(\frac{1-{m}}{1+{m}}\right)^{\half}.
$$
which yields 
$$
a  = \frac{1}{2} (x+(x^{-1})^*) = (1-{m}^2)^{- \frac{1}{2}}
$$
and
$$
b = \frac{1}{2} (x-(x^{-1})^*) =  -{m}(1-{m}^2)^{- \frac{1}{2}}.
$$
Conversely, if 
the element  $g\in {\mu}(A)$ is determined by matrix 
entries $a$ and $b$ of the above form,
then
$$
x=a+b= (1-{m}^2)^{-\half} - {m}(1-{m}^2)^{-\half}= \left(\frac{1-{m}}{1+{m}}\right)^{\half}, 
$$
which shows that $x$ is positive.
\qed

There is an alternative parametrization of
the positive elements
of the \mobius group ${\mu}(A)$.
\bproposition
 An 
$g \in {\mu}(A)$ is positive if and only if
it can be written as
$$
g=
\left(
\begin{array}{cc}
{\cosh {\omega}} & {\sinh {\omega}}\\
{\sinh {\omega}} & {\cosh {\omega}}
\end{array}
\right),
$$
for a selfadjoint operator ${\omega}$ in $A$.
This ${\omega}$ is uniquely determined by $g$.
\eproposition
\proof
Let  a positive element $x$ of  $GL_1(A)$
be  the image of $g\in \mu(A)$. 
 For a positive $x \in 
GL_1(A)$ there exists a unique bounded self-adjoint 
element $\omega$ such that 
$x =e^{\omega}$. The elements $g = \larray a & b \\ b & a \rarray$
is reconstructed from $x$ using the identities
$a=\half (x+(x^*)^{-1})= {\cosh {\omega}}$ and
$b=\half (x-(x^*)^{-1})= {\sinh {\omega}}$
which gives the proof of Proposition. \qed

An arbitrary bounded invertible operator $A$
in a complex Hilbert space ${\cal{H}}$
admits two {\em polar decompositions},
i.e., it can be written either as $A=U_L(A^*A)^{\half}$,
or as $A=(AA^*)^{\half}U_R$ with $U_L$
and $U_R$ unitary operators.
Both decompositions are of course unique.
Through the isomorphism between $GL_1({A})$
and ${\mu}({A})$
an arbitrary element $g$ of ${\mu}({A})$
has a corresponding pair
of what may also be called polar decompositions.

\bproposition
 Every  element
$g$ of the \mobius group ${\mu}({A})$
can be written uniquely as a product of a unitary element and
a positive one. In other words, 
\beqn\label{polar}
\mbox{\rule{5mm}{0mm}} g= \left(
\begin{array}{cc}
a & b\\
b & a
\end{array}
\right)
=
\left(
\begin{array}{cc}
u & 0\\
0 & u
\end{array}
\right)
\left(
\begin{array}{cc}
(1-{m}^2)^{-\frac{1}{2}} & -{m}(1-{m}^2)^{-\frac{1}{2}}\\
-{m}(1-{m}^2)^{-\frac{1}{2}} & (1-{m}^2)^{-\frac{1}{2}}
\end{array}
\right).
\eeqn
In this $u=a(a^*a)^{-\frac{1}{2}}$
and ${m}=-a^{-1}b$.

\medskip
\noindent
Similarly $g$ can be written as
$$
g=
\left(
\begin{array}{cc}
(1-{m}^2)^{-\frac{1}{2}} & -{m}(1-{m}^2)^{-\frac{1}{2}}\\
-{m}(1-{m}^2)^{-\frac{1}{2}} & (1-{m}^2)^{-\frac{1}{2}}
\end{array}
\right)
\left(
\begin{array}{cc}
u & 0\\
0 & u
\end{array}
\right),
$$
where now $u=(aa^*)^{-\frac{1}{2}}a$
and ${m}=-ba^{-1}$.
\eproposition
\proof The statement  follows directly from Proposition \ref{unitary-and-positive}. \qed

This result was first proved by Connes \cite[Prop. 5, p. ~335]{Connes:Book}. 

Let us denote by $\calu$ the group of unitary elements in $GL_1(A)$. 
\bproposition\label{retract}
The unitary group $\cal{U}$ is a deformation retract of the group $\mu(A)$.
\eproposition
\proof  
Let us denote by $g_m$ the matrix
$$
g_m = \left(
\begin{array}{cc}
(1-{m}^2)^{-\frac{1}{2}} & -{m}(1-{m}^2)^{-\frac{1}{2}}\\
-{m}(1-{m}^2)^{-\frac{1}{2}} & (1-{m}^2)^{-\frac{1}{2}}
\end{array}
\right)
$$
so that $g = ug_m$ in \ref{polar}. 
Replacing $m$ by $tm$, $t\in [0,1]$,   we obtain a 
continuous path of elements $g_{m,t}$ linking $g_m$ to the identity matrix. 
\qed 

\bremark
In what follows, we shall be mostly interested in the 
case where $A$ is identified with a $*$-subalgebra of the 
 algebra  of bounded operators on a Hilbert space by 
means of a faithful representation $\pi$. The \mobius group
of interest in that case will  be constructed in the same way as above
with the assumption that the algebra $A$ is the commutant $M$ of 
the representation $\pi$. Our results will still apply in this case. 
The group $\mu(M)$ (or $\mu_{ev}(M)$ in the graded case) is 
the same as the group introduced by Connes in \cite[p.334]{Connes:Book}. 
It would be natural to call this group the {\em Connes-\mobius} group, 
but for reasons of simplicity we shall use the name \mobius group.
\eremark

\mysection{\mobius transformations of Fredholm modules}

\mysubsection{Fredholm modules}

Probably the most interesting feature of the \mobius 
group is the fact that it acts on the space of Fredholm 
modules. This action, as we shall see later, 
has interesting geometric consequences. But first, 
let us quickly review the standard notions. 

\bdefinition\label{definition}
Let $A$ be an involutive algebra over $\bc$. An {\em even} Fredholm module 
over $A$ is given by the data  $( \bh, \pi, \gamma,  F)$.
Here 
 $\bh $ is a $\bz_2$-graded Hilbert space with the grading 
given by a 
self-adjoint involution $\gamma$; $\pi$ is a
faithful $*$-representation $\pi$ of $A$ in $\bh$ commuting 
with $\gamma$, and $F$ is a self-adjoint involution  on $\bh$ which 
anticommutes with 
$\gamma$. Moreover, we assume that for any $a\in A$, $[F, \pi(a)]$ 
is a compact operator.

An {\em odd} Fredholm module is defined in an analogous way but without the 
assumption that  the space $\bh$ is $\bz_2$-graded. 
\edefinition
 
 \bremark
The presence of a self-adjoint involution $F$ anticommuting with 
$\gamma $ guarantees that the eigenspaces $\bh^0$ and $\bh ^1$ are isomorphic.
Conversely, any unitary operator $T : \bh ^0 \ra \bh ^1$ gives rise to 
a self-adjoint  involution $F$ given by the formula
$$
F = \larray 0 & T^* \\ T & 0 \rarray
$$
Using this idea it is not difficult to show that the space of all s.a. 
involutions $F$ anticommuting with $\gamma$ is diffeomorphic to the group 
$U(\bh^0)$ of unitary operators on $\bh^0$.
\eremark

\bdefinition
Two  Fredholm modules $\alpha_j= (\bh_j , \pi_j, F_j, \gamma_j)$, $j=1,2$, 
are called {\em unitarily equivalent} 
if and only if there  exists a unitary map $U : \bh_1 \ra \bh_2$ such that
$$
F_2=UF_1U^{-1} , \qquad\gamma_2= U\gamma_1 U^{-1}, 
$$
and
$$
U\pi_1(a) U^{-1} = \pi_2(a), \forall a \in A. 
$$
which will be  written as $U\pi_1 U^{-1} = \pi_2$. 

In such a case we shall write $\alpha_2 = U\alpha_1 U^{-1}$. 
\edefinition

One can also  define a more general notion of isomorphism of Fredholm 
modules, but we shall not need it here. 

It is not difficult to understand the structure
of the space of unitarily equivalent Fredholm modules.  
Let us denote by ${\rm FM}$ the class of all odd Fredholm modules 
$\alpha =  (\bh , \pi, F)$. In the even case
we shall use the symbol ${\rm FM}^{ev}$ to denote the class of 
all even Fredholm modules $\alpha = (\bh , \pi, \gamma, F)$. 

Two Fredholm modules $\alpha $ and $\alpha'$ can only be 
unitarily equivalent if the representations $\pi$ and $\pi'$ 
belong to the same unitary conjugacy class. Thus we have 
the following disjoint union decomposition 
$$
\fm = \bigcup_p\fm_p
$$
where $p$ runs through the space of equivalence classes of representations 
$\pi$ and $\fm_p$ is the set of all
Fredholm modules $(\bh , \pi, F)$ in which $\pi\in p$. 
If we denote the relation of unitary equivalence of 
Fredholm modules by $\sim$ then we have the following simple
observation: 
\beqn\label{equivalence1}
\fm/\sim\; = \;\union_p\left(\fm_p/\sim\right)
\eeqn
In the even case we have in the same way 
$$
\fm^{ev}/\sim\; = \;\union_p\left(\fm^{ev}_p/\sim\right)
$$
Since we only consider separable Hilbert spaces, we can describe
the set of equivalence classes $\fm_p/\sim$ ($\fm_p^{ev}/\sim$, respectively)
in a simple way. Since the reasoning is the same in the even and odd
cases, we shall concentrate on the odd case. 
Let us fix a Hilbert space $\bh_p$ equipped
with a representation $\pi_p\in p$. We define $\fm_{\pi_p}\subset \fm_p $ 
to be the set of Fredholm modules $\alpha= ( \bh_p, \pi_p, F)$, 
where now only $F$ is variable. We remark that if $\pi, \rho \in p$
then $\fm_{\pi_p}\simeq \fm_{\rho_p}$.

Let us for the moment  denote by $\sim_p$ the restriction of 
the equivalence relation $\sim$ to the set $\fm_{\pi_p}$. 
It follows directly from the definition 
that a unitary equivalence of two Fredholm modules from
$\fm_{\pi_p}$ is established by means of a unitary operator
$U$ from the {\em commutant} of the fixed representation $\pi_p$.

For each $\alpha =(\bh_\alpha, \pi_\alpha,  F_\alpha) \in \fm_p$ one can find a Fredholm module $\beta\in \fm_{\pi_p}$ such that $\alpha\sim\beta$. Indeed, as $\pi_\alpha $ and 
$\pi_p$ are unitarily equivalent representations, 
there exists a unitary isomorphism $U: \bh_\alpha \ra \bh_{\pi_p}$
such that $\pi_p = U\pi_\alpha U^{-1}$. Define $\beta 
= (\bh_p, \pi_p, UF_\alpha U^{-1})$. It
is clear that $\beta \in \fm_{\pi_p}$ and $\alpha \sim \beta$. 
Hence, using general properties of equivalence relations, 
we conclude that there exists a bijection between the 
sets of equivalence classes $\fm_p /\sim$ and $\fm _{\pi_p}/\sim_p$. 

Repeating this construction in the even case we arrive at the 
following \lq localization' result. 

\btheorem\label{local}
There are the following bijections of sets: 
$$
(\fm/\sim)\; \simeq \; \union_p (\fm_{\pi_p}/\sim_p)
$$
$$
(\fm^{ev}/\sim)\; \simeq \; \union_p (\fm^{ev}_{\pi_p}/\sim_p)
$$
where $p$ runs through the set of conjugacy classes of representations
of $A$ in $\bh$ and $\pi_p$ is a fixed representative of a given class $p$. 
\etheorem

In other words, if we want to find all Fredholm modules unitarily 
equivalent to a given $\alpha = ( \bh , \pi, \gamma, F)$, it is sufficient 
to consider, without losing any information, the Fredholm modules that 
can be obtained by acting on $\alpha$ with unitary operators that
belong to the commutant of $\pi$.

\mbox{}

We pass  to the discussion of the action of the \mobius 
group on the space of Fredholm modules. This action was 
first introduced by Connes, and is defined as follows. 
If $\alpha = (\bh, \pi,
F, \gamma)$ is an even Fredholm module and $g = \larray a & b \\ b & a 
\rarray$, where $a,b\in M= (\pi(A))'$ is an element of the \mobius group
$\mu(M)$, then $g\alpha=
(\bh, F', \gamma)$ where 
$F' = (a F + b ) ( bf - a)^{-1}$. As Connes shows, the \mobius group 
maps $p$-summable Fredholm modules to $p$-summable Fredholm modules of 
the same parity. There is an interesting generalization of this statement.

 For any two bounded operators $P,Q$ on $\bh$
we write $P\sim Q$ if $P$
and $Q$ differ by a compact operator.

\bdefinition
The essential commutant of  a representation $\pi$ in $\call$ is
$$
D_\pi(A) = \{ x\in \call \mid \forall a \in A,\; \;  [\pi(a), x]\sim 0\}
$$
\edefinition
$D_\pi(A)$ is a $C^*$-subalgebra of the algebra $\call$ of bounded operators 
on $\bh$. It is sometimes called the dual algebra of $A$, associated to the 
representation $\pi$. 
The importance of the algebra $D_\pi(A)$ lies in its close relation to the 
$K$-homology \cite{Higson}. 

Let $A$ be a $C^*$-algebra,  let $\pi$ be its representation in a 
separable Hilbert space $\bh$, and let 
$(A, F, \bh)$ be a Fredholm module over $A$. We shall denote by 
$D = D_\pi(A)$ the
essential commutant of $\pi(A)$ in $\call$. Associated with this 
data is the \mobius group $\mu(D)$. 
 We have seen that $\mu(D)$ sends self-adjoint involutions to self-adjoint
 involutions, and so in order to understand its action on the 
 space of Fredholm modules we need to understand its behaviour 
 with respect to the compactness condition. First we treat the 
 case of a Fredholm module, where we assume that
 the commutator $[F,x]$ is a compact operator for any element $x$ of 
 the algebra $A$. 
\bproposition\label{compact-commutator}
Let $[F,x]$ be a compact operator for all $x\in A$. Then the same
is true for $F' = g(F)$, where $g\in \mu(D)$. In the case of an 
even Fredholm module, this statement holds for any $g\in \mu_{ev}(D)$. 
\eproposition
\proof We simply need to calculate the commutator $[F', x]$ in 
terms of $[F,x]$. We have 
$$
\begin{array}{l}
[(aF + b)(bF + a)^{-1} , x]   \\
\rule{10mm}{0mm}=[aF + b, x](bF+a)^{-1} + 
(aF+b)[(bF+a)^{-1}, x]
\end{array}
$$
The first term in this sum may be written as 
\beqn\label{first}
[aF + b, x](bF+a)^{-1}  = (a[F,x] + [a,x]F+ [b,x])(bF+a)^{-1}
\eeqn
whereas the second takes the form
\beqn\label{second}
\begin{array}{l}
(aF+b)[(bF+a)^{-1}, x]  \\ 
\rule{10mm}{0mm} = - (aF+b)(bF+a)^{-1}(
b[F,x] + [b,x]F + 
[a,x])(bF+a)^{-1}
\end{array}
\eeqn
The first terms of formulae \ref{first} and \ref{second}
combine to give 
$$
\begin{array}{c}
a[F,x] (bF+ a)^{-1} - (aF + b)(bF+a)^{-1} b[F,x](bF+a)^{-1} \\
\rule{45ex}{0mm}
= ((bF + a)^{-1})^*[F,x](bF +a)^{-1}
\end{array}
$$
if we use that $(a-F'b)^{-1} = (bF + a)^*$. The remaining terms
all contain either $[a,x]$ or $[b,x]$ and so their sum is a compact
operator. To summarize, we have
\beqn\label{transform-of-commutators}
[F', x] = ((bF + a)^{-1})^*[F,x](bF +a)^{-1} + K
\eeqn
where $K$ is a compact operator. \qed

\bremark
Apart from the essential commutant $D_\pi(A)$, we can define
a $p$-summable commutant 
$$
D^p_\pi(A)= \{ x\mid \forall a \in A, \; \; [x,\pi(a) ]\in \call^p(\bh)\}
$$
This is still a subalgebra of $\call$, but it  need not be a $C^*$-subalgebra, 
as the Schatten ideals $\call^p$ are not norm-closed in 
$\call$. This problem disappears if we restrict our attention to the 
{\em commutant} $M$ of $\pi(A)$, which brings us back to the situation 
considered by 
Connes \cite[p.~335]{Connes:Book}. 
In this case Lemma \ref{compact-commutator} can be improved: 
The group $\mu(M)$ maps  $p$-summable Fredholm modules 
to $p$-summable Fredholm modules, for any $p\leq 1$. In the even case, 
the group $\mu_{ev}(M)$ acts in the same way on even 
Fredholm modules.  
\eremark

\mysubsection{\mobius group and $K$-homology}

The idea  that gave rise to the notion of  Fredholm module can 
be traced back to the important paper \cite{Atiyah} of Atiyah, where 
he describes an abstract approach to the calculus of elliptic 
operators on a topological space in his search for a theory 
that could be considered dual to the $K$-theory.  This idea was 
developed in papers by, among others, Baum, Connes, Douglas, Higson, 
Kasparov and led to a definition of $K$-homology of a $C^*$-algebra $A$ (cf. \cite{BD}\cite{Higson}). 

The starting point is the space of all Fredholm modules over a 
$C^*$-algebra $A$, represented in a fixed Hilbert space $\bh$. 
On this space one defined the notion of equivalence using the already
discussed unitary equivalence, triviality and homotopy of Fredolm modules.
Let us introduce the latter two concepts. 
 
An even Fredholm module 
$\alpha = ( \bh , \pi, F, \gamma)$ 
is trivial if and only if $[\pi (a), F]= 0$ for all $a\in A$.

Two Fredholm modules $\alpha$ and $\beta$ are said to be
homotopic if there exists a family $(\bh, \pi , \gamma,  F_t)$, 
where $t\in [0,1]$ and $F_t$ varies norm-continuously in the space of self-adjoint 
involutions anticommuting with $\gamma$. We keep the representation $\pi$ fixed here. 
These definitions are modified in the obvious way to the case of odd 
Fredholm modules.

Using the notions of unitary equivalence, triviality and homotopy 
equivalence we define an equivalence relation on the space 
of Fredholm modules. The space 
of equivalence classes of even Fredholm modules with respect to this relation
 gives rise to the even $K$-homology group $K^0(A)$. Similarly, the 
same idea applied to the space of odd Fredholm modules produces
the group $K^1(A)$. Since the \mobius group acts  
on Fredholm modules, it is natural to ask about the behaviour 
of this action from the point of view of $K$-homology. 

It is clear that the \mobius group maps trivial Fredholm modules to 
trivial Fredholm modules. Moreover, we have the following result.  
\bproposition
The \mobius group $\mu(A)$ ($\mu_{ev}(A)$) in 
the even case) acts on the space of homotopy classes of Fredholm modules. 
\eproposition
\proof 
If $F_t$ is a norm-continuous family of involutions between $F_0$ and $F_1$, 
then 
$$
g(F_t) = (aF_t + b)(bF_t+a)^{-1}
$$
is a norm-continuous path of self-adjoint involutions between $g(F_0)$ and 
$g(F_1)$. \qed

\btheorem
The \mobius group $\mu(M)$ acts trivially on the $K$-homology of the 
algebra $A$. 
\etheorem
\proof 
Let $\alpha = (\bh ,\pi,  F)$ be a Fredholm module and let $g\in \mu(M)$, 
where $M$ is the commutant of the representation $\pi(A)$. The \mobius 
transformation $g$ maps $\alpha $ to the Fredholm module 
$g(\alpha) = ( \bh , g(F), \pi) $. Assume that the polar decomposition 
of $g$ is $g = ug_m$. Then by Proposition \ref{retract}, $g(\alpha)$ 
is homotopic to the Fredholm module $u\alpha u^*$, which belongs to the 
same $K$-homology class as $\alpha$. In the even case, an even  \mobius 
transformation
$g$ maps an 
even Fredholm module $\beta = (\bh , \pi, F, \gamma)$ to 
$g(\beta) = (\bh , \pi, g(F), g(\gamma))$, which is homotopic to the 
even Fredholm module
$$u\beta u^* = (\bh, \pi, uFu^*, u\gamma u^*)=
(\bh , \pi , uFu^* , \gamma),$$
 which 
belongs to the same $K$-homology class as $\beta$. \qed

\mysubsection{Polarized modules}

To define a Fredholm module requires that we specify a Hilbert space
$\bh$. In many sitations, however, the Hilbert  space structure 
is not really relevant, and what really counts is just the underlying 
locally convex topological space. Of course, we cannot hope that 
an arbitrary locally convex topological space could be used 
to gain useful information about geometric objects. There
is, however, a category of spaces which are equipped 
with indefinite inner products, whose properties make them useful
in geometric situations. These are {\em Krein spaces}.

Let us recall that a {\em Hilbertian space} $\calh$ is a locally convex 
topological vector space which is isomorphic to a Hilbert space. In 
other words, there exists an inner product on $\calh$ which induces
the same topology.
\bdefinition
 A {\em Krein space} is a complex Hilbertian space
${\cal{H}}$,
provided with a continuous hermitian form
$$
\sigma : {\cal{H}} \times {\cal{H}} \rightarrow C
$$
and such that there exists at least one
compatible Hilbert space inner product $(\cdot , \cdot )$
on ${\cal{H}}$ given by
$$
(x,y) = \sigma( x, {\gamma}y)    ,
$$
for all $x$ and $y$ in ${\cal{H}}$,
with ${\gamma}$ an involution operator,
selfadjoint with respect to $\sigma(   \cdot, \cdot)    $.
\edefinition
One verifies easily that the  following three statements are equivalent:
\begin{enumerate}
\item $( \cdot , \cdot )$ is a hermitian form.
\item  $\sigma(  {\gamma}x, y )     = \sigma(  x, {\gamma}y)    , \forall x,y \in {\cal{H}}$.
\item  $({\gamma}x, y ) = (x, {\gamma}y), \forall x,y \in {\cal{H}}$.
\end{enumerate}

A compatible inner product is one that equipes the Hilbertian space
$\calh$ with a Hilbert space structure whose topology,  
determined by the inner product, is the same as the original locally 
convex topology of $\calh$. In what follows, we shall frequently 
need to consider the same topological space either as a Krein space, 
equipped with an indefinite inner product $\sigma( \cdot , \cdot )$
or as a Hilbert space, with a Hilbert space inner product $(\cdot , \cdot)$. 
To distinguish between these two cases, we shall use the notation $\calh$
for Krein spaces and $\bh$ for Hilbert spaces. 

\bdefinition
 A {\em polarized module} over a unital $*$-algebra
${\cal{A}}$ is a triple
${\calp}=({\cal{H}}, {\pi}, {\cal{E}})$
in which one has
\begin{enumerate}
\item  a Krein space $({\cal{H}}, \sigma(   \cdot , \cdot )     )$;
\item  a faithful unitarizable $*$-representation ${\pi}$
of ${\cal{A}}$ by continuous linear operators
in ${\cal{H}}$;
\item  a closed linear subspace ${\cal{E}}$
of ${\cal{H}}$, on which the hermitian form
$\sigma(   \cdot , \cdot )    $ vanishes.
\end{enumerate}
We assume, moreover, that 
\begin{enumerate}
\setcounter{enumi}{3}
\item the operator $T: \calh \ra \calh^*$ associated 
with the bilinear form $\sigma( \cdot , \cdot ) $ maps 
the subspace $\cale $ onto its annihilator $\cale^{an}$
in the topological dual $\calh^*$ of the space $\calh$; 
\item  the operator $T_a: \cale \ra \cale^*$
from $\cale$ to its topological dual given by 
$\cale \ni x \mapsto \sigma( a\cdot , x )$, 
 is compact for any $a\in \cala$. 
\end{enumerate}
\edefinition
\bremark
It would be more appropriate to use the name {\em polarizable} 
module, but in the interest of simplicity we shall use our 
current convention. This notion was first introduced in a slightly 
different form in \cite{BCE}.

We remark also that a representation is unitarizable iff
there exists a Krein involution $\gamma$ such that, in the corresponding 
Hilbert space $\bh_\gamma$, the representation $\pi$ 
becomes a unitary representation. A sufficient condition for this 
to hold is the existence of a Krein involution $\gamma$ commuting with $\pi$. 

The problem of existence of unitarizable representations of $A$ is
closely related to the similarity problem for a representation 
of a $C^*$-algebra, which was solved in several special cases
\cite{H}\cite{Ch}.

Our final remark on the definition of the polarized module is that
although it may seem unnatural to single out a fixed subspace 
of $\calh$, there are examples in which such a subspace 
appears quite naturally. For instance, in our most important case
of an even dimensional smooth manifold, the Hilbertian space $\calh$ 
will be the 
locally convex space of differential forms in the middle dimension. 
Then $\cale$ is the closed subspace generated by the image of 
the de Rham differential. 
\eremark

One can introduce the  notion of unitary equivalence of polarized modules
which is analogous to that defined in the case of Fredholm modules. 
We state it here for the convenience of the reader. 
\bdefinition
Two polarized modules ${\calp _i} = (\calh_i , \pi_i, \cale_i)$, $i=1,2$
are unitarily equivalent iff there exists a Krein-unitary map
(i.e., 
unitary with respect to the bilinear form $\sigma(\cdot, \cdot)$) 
$U:\calh_1 \ra \calh_2 $ such that 
$$
U(\cale_1) = \cale_2, \qquad {\pi}_2 = U{\pi}_1U^{-1}
$$ 
\edefinition

Polarized modules are more fundamental objects than Fredholm modules
as we now set out to explain. This relation will be important when 
we discuss the geometric example of a compact conformal manifold. 
Our exposition here is a review and an extension of results from \cite{BCE}. 
\bproposition\label{Fredholm-to-polarized}
To any even Fredholm module $\alpha =  ({\cal{H}}, {\gamma}, {\pi}, F)$
one can associate a unique polarized module. 
\eproposition
\proof
We put
$$
\sigma(   x, y)     = (x, {\gamma}y), \, \forall x,y \in {\cal{H}}.
$$
This is a continuous hermitian form,
which is indefinite and nondegenerate. The space 
${\cal{H}}$ equipped with the form $\sigma( \cdot, \cdot )$
becomes a Krein space.

Let $\cale$ be the $+1$-eigenspace of the involution $F$. Then 
${\cal{E}}$ is a closed linear subspace of ${\cal{H}}$.
Moreover, for $x$ and $y$ in ${\cal{E}}$ one has
$$
\sigma(   x, y )     = (x, {\gamma}y) = (Fx,
 {\gamma}Fy) = -(Fx, F{\gamma}y) = - \sigma(  x,y )    ,
$$
which implies $\sigma(  x, y )     = 0$ for all $x$ and $y$ in ${\cal{E}}$, and
so the subspace $\cale$ is totally isotropic, as required.

Since the Hilbert space $\bh$ comes equipped with the orthogonal 
sum decomposition $\bh = \cale \oplus \gamma(\cale)$, we see that
the (Hilbert space) orthogonal complement $\cale^\perp$ of $\cale$
is  $\cale^\perp = \gamma (\cale)$. Using this fact together 
with the Riesz representation theorem we conclude that the 
annihilator $\cale^{an}$ of the subspace $\cale$ is conjugate-isomorphic
with $\gamma(\cale)$. In other words, for any $\xi \in \cale^{an}$
there exists a unique $y = \gamma x \in \gamma(\cale)$, where $x\in \cale$
such that 
$$
\xi = ( \cdot, y) = (\cdot , \gamma x ) = \sigma( \cdot , x ). 
$$
This shows that the map $T: x\mapsto \sigma( \cdot , x )$
from $\calh$ to dual
maps $\cale$ onto $\cale^{an}$. 

It remains to check the compactness condition. For any $a\in A$ and 
any $x', y' \in \cale$ we have
$$
(ax', \gamma y' ) = (aPx, \gamma Py)
$$
where $x,y\in \bh$ and $P = (1+F)/2$ is the orthogonal projection onto 
$\cale$. So the compactness of the operator $T_a$ 
 will be proved if we check 
that the operator $P\gamma a P $ is compact for any $a\in \cala$. We have
$$
\begin{array}{rcl}
P\gamma a P & = & \gamma (1-F)a(1+F)/4\\
& = & \gamma ([F, a] F -[F,a])/4
\end{array}
$$
and so the result follows from the compactness condition satisfied 
by the Fredholm module.

This way we have defined a polarized module
${\calp}(\alpha) = ({\cal{H}}, {\pi}, {\cal{E}})$
underlying  the Fredholm module $\alpha$. 
\qed
\bproposition
 Suppose that two Fredholm modules $\alpha_1$
and $\alpha_2$ are unitarily equivalent.
Then the same is true of the underlying polarized  modules 
${\calp}(\alpha_1)$ and
${\calp}(\alpha_2)$.
Moreover, the even unitary map $U$
$$
U \, : \, {\cal{H}}_1 \, \rightarrow \, {\cal{H}}_2
$$
intertwining  the representations ${\pi}_1$ and ${\pi}_2$
and  the operators $F_1$ and $F_2$,
becomes a Krein unitary map, mapping the subspace ${\cal{E}}_1$ onto 
the subspace ${\cal{E}}_2$.
\eproposition
{\em Proof}\,: Let $\alpha_j = (\bh_j,  {\pi}_j,
F_j)$,
for $j =1,2$, be two unitarily equivalent Fredholm modules.
One defines again the corresponding polarized modules ${\calp}_j = 
{\calp}(\alpha_j)$, for $j =1,2$, as described above.
The unitary equivalence of $\alpha_1$ with  $\alpha_2$ is established by means 
of a map
$U: {\bh}_1 \rightarrow {\bh}_2$,
which has the property $F_2=U F_1U^{-1}$.

It remains to prove that  ${\cal{E}}_2$
is the image of ${\cal{E}}_1$ under the map $U$.
For this one observes that $x \in {\cal{E}}_1$ is a $+1$-eigenvector of
$F_1$, i.e. $F_1x=x$ which is the same as
$U^{-1}F_2Ux=x$. It follows that 
$F_2Ux=Ux$, which means that  $Ux \in {\cal{E}}_2$. 
From this it follows that ${\cal{E}}_2=U({\cal{E}}_1)$.
The last  statement about the map $U$ is  clear.
\qed

It is an important fact that a polarized module can be lifted to 
an even Fredholm module. This lift is nonunique, in general, and 
the freedom of choice  involved  here is conveniently described by the action 
of the \mobius group, as we shall see shortly.
\bdefinition
 An involution operator ${\gamma}$ in the Krein space
${\cal{H}}$ of the polarized module
${\calp} = ({\cal{H}}, {\pi}, {\cal{E}})$
is called {\em compatible}
iff it is a Krein-self ajdoint involution commuting with the 
representation $\pi$,  which makes $\calh$ into  a Hilbert
space equipped with the inner product 
$(\cdot, \cdot )_\gamma = \sigma( \cdot,\gamma  \cdot ) $. 
The set of all compatible involutions associated with a polarized 
module $\calp$
will be denoted ${\Gamma}_{\cal P}$.
\edefinition
We remark that the space of compatible involutions of a given polarized
module is nonempty. 
\bproposition\label{Cayley-transform}
Let $\gamma_0$ and $\gamma_1$ be two compatible involutions. Then 
$$
\gamma_1 = (1+m )\gamma_0 (1 +m)^{-1}
$$
where $m$ is a uniquely determined operator which is self-adjoint
with respect to the scalar product $(\cdot, \cdot)_0$ and such 
that $\norm{m}_0<1$. Moreover, $m$ anticommutes with $\gamma_0$ 
and belongs to the commutant $M$ of the representation $\pi$. 
\eproposition
\proof Let $Q = \gamma_0\gamma_1$. Then $Q$ is a bounded invertible
operator, which is positive with respect to the inner product $(\cdot , 
\cdot )_0$. 
To check the last statement we simply note that 
$$
(x , Qx)_0 = (x, \gamma_0\gamma_1 x)_0 = \sigma( x, \gamma_1x )
= (x,x)_1 \geq 0
$$

We shall now use the Cayley transform, which establishes
a 1-1 correspondence between bounded, positive, invertible 
operators $Q$ and bounded self adjoint operators $m$ of norm 
strictly smaller than $1$. More precisely, for each operator
$Q$  there exists a unique operator $m$
$$
Q = \frac{1-m}{1+m}
$$
with the inverse relation given by the formula
$$
m = \frac{1-Q}{1+Q}.
$$
Thus for a unique $m$, 
$$
Q= \gamma_0\gamma_1= \frac{1-m}{1+m}
$$ 
and so 
$$
\gamma_1 = (1+m )\gamma_0 (1 +m)^{-1}
$$

To check that $m$ anticommutes with $\gamma_0$, 
we use that 
$$
( 1+Q)\gamma_0(1-Q) + (1-Q)\gamma_0(1+Q) = 0
$$
which after multiplying on the left by $(1+Q)^{-1}$ gives
$$
\gamma_0  \frac{1-Q}{1+Q} + \frac{1-Q}{1+Q}\gamma_0 = 
\gamma_0 m + m\gamma_0 = 0
$$
as required. 

To show that $m$ commutes with $\pi(a)$, for all $a\in A$, 
we note that $Q$ has this property, and then use the 
Cayley transform together with the continuous functional 
calculus to conclude that the same is true for $m$. 
\qed

\bproposition
A polarized module $\calp = (\calh , \pi, \cale)$ can be lifted 
to a Fredholm module $\alpha (\calp) = (\bh , \pi, \gamma, F)$, 
where $\gamma \in \Gamma_{\cal P}$. 
\eproposition
\proof Let us choose a compatible involution $\gamma \in \Gamma_{\cal P}$. 
It turns the Krein space $\calh$ into a Hilbert space 
$\bh_\gamma$ equipped with the inner product $(\cdot , \cdot )_\gamma = 
\sigma(  \cdot , \gamma \cdot )$. The Hilbert space $\bh_\gamma$
is equipped with the orthogonal direct sum decomposition $\bh_\gamma
= \cale \oplus \gamma(\cale)$. Define $F$ to be $+1$ on $\cale$ and 
$-1$ on $\gamma(\cale)$. It is clear that $F$ is an involution anticommuting
with $\gamma$ and that it is self-adjoint with respect to the 
inner product $(\cdot , \cdot )_\gamma$. 

We need to check that the commutator $[F,a]$ is a compact operator for any 
$a\in A$. To this end we need to use the assumption that the operator
$$
T_a : \cale \ra \cale^*; \quad x \mapsto \sigma( a\cdot , x )
$$
is compact for every $a\in A$. Then also the operator 
$x \ra (a\cdot , \gamma x )_\gamma$ is compact, which in turns shows that 
the operator $P\gamma a P$, where $P = (1+F)/2 $ is the orthogonal projection 
onto $\cale$, is also compact for any $a\in A$. Multiplying on the left by $\gamma$ we see that the operator
$$
\gamma P \gamma aP = (1-P)a P
$$
is compact, if we use that $\gamma P = (1-P)\gamma$. Since the above equality
holds for all $a$ in $A$, taking the adjoints we see that the operator
$Pa(1-P)$ is also compact for all $a\in A$. Adding the two together
we get that 
$$
(1-P)aP + Pa(1-P) = [P,a] = [F,a]/2
$$
is compact. This gives the proof of the result. 
\qed

\mysubsection{Lifting polarized modules} 

What happens if we lift a given polarized  module $\calp = (\calh , \cale, \pi)$ in two 
different ways, by choosing two different compatible involutions $\gamma_1$ and $\gamma_2$ 
in $\Gamma_{\cal P}$? 
 
The two choices give rise to two Fredholm modules $\alpha_j = (\bh_j, \pi, \gamma_j, F_j)$, 
$j= 1,2$, 
where $\bh_j$ is the Hilbert equipped with the inner product $(\cdot, \cdot )_j 
= \sigma (\cdot , \gamma_j\cdot)$, and  $F_j$ is the involution that is $+1$ on 
$\cale$ and $-1$ on $\gamma(\cale)$. We are going to construct a map between 
these Fredholm modules in a few stages. 

First we note that  the two 
Hilbert space inner products are related in the following 
way.  Let $x,y\in \calh$. Then 
\beqn\label{comparing-inner-products}
(x,y)_1 = \sigma (x, \gamma_1 y) = \sigma( x , \gamma_2^2\gamma_1 y) = 
(x,\gamma_2\gamma_1 y)_2.
\eeqn
Let $Q=\gamma_2\gamma_1$. Then $Q$ is an invertible operator, which 
is self-adjoint with respect to the inner product 
$(\cdot , \cdot )_1$ as can be seen from the following simple
calculation:
$$
(x, Qy)_1 = \sigma (x, \gamma_0 y) = \sigma (\gamma_0 x, y) 
= (\gamma_0 x, \gamma_1y)_1 = (\gamma_1\gamma_0 x, y)_1 = (Qx, y)_1. 
$$
Moreover, our assumptions about the form $\sigma$ imply that $Q$ is 
a positive operator. 
Let us denote by $W$ the square root of $Q$. Then, by \ref{comparing-inner-products}, we 
may regard $W$ as a unitary operator 
$W: \bh_0\ra \bh_1$.

Let $g_{21}$
be the following matrix
\beqn \label{gen-mobius}
g_{21} = \larray \cala & \calb \\
\calb & \cala \rarray = 
\larray \half(1+Q) & \half(1 - Q)\\
\rule{0mm}{5mm} \half(1 - Q)& \half(1+ Q)
\rarray
\eeqn
where $\cala$ and $\calb$ are regarded as operators from $\bh_1\ra \bh_2$. 
\bproposition
\begin{enumerate}
\item $\cala$ and $\calb$ commute with the representation $\pi$. 
\item The operators $\cala$ and $\calb$ satisfy  Connes identities \ref{Connes-group}. 
\item The matrix $g_{21}$ is invertible and its inverse is given by 
$$
g_{21}^{-1} = \larray \cala^* & - \calb^* \\
-\calb^* & \cala^*
\rarray .
$$
\item $g_{21}$ admits the following \lq right' polar decomposition (with analogous
formula holding for the \lq left' polar decomposition)
\beqn\label{new-polar}
g_{21} = \left( 
\begin{array}{cc}
(1-m^2)^{- \half} & m(1-m^2)^{-\half}\\
 m(1-m^2)^{-\half} & (1-m^2)^{-\half}
\end{array}
\right)\larray W & 0 \\ 0 & W\rarray
\eeqn
where $m $ is the Cayley  transform of $Q$, $m = (1- Q)(1+Q)^{-1}$
and $W = Q^\half$.
\end{enumerate}
\eproposition
\proof 
The first statement of the Proposition is obvious given that all compatible
involutions commute with $\pi$. The second statement, which is checked by a simple calculation, 
directly 
implies the third. Finally, the proof of the polar decomposition 
\ref{new-polar} requires only a small modification of the proof of \ref{polar}.
\qed

We remark  that the first factor on the right in  formula  \ref{new-polar}
is 
just a positive \mobius transformation defined relative to the commutant of the
representation $\pi$ in the Hilbert space $\bh_2$, whereas the factor on 
the right is simply a unitary isomorphism identifying the space $\bh_1$ with $\bh_2$. 
\bdefinition
We shall call the matrix $g_{21}$ a {\em generalized \mobius transformation}
associated with the ordered pair $(\gamma_2, \gamma_1)$ of compatible involutions
from $\Gamma_{\cal P}$. 
\edefinition
The following result is proved by an easy calculation.
\bproposition\label{groupoid}
Let $\gamma_1$, $\gamma_2$ and $\gamma_3$ be compatible involutions. Then 
$$
g_{31} = g_{32}g_{21}.
$$
\eproposition

Thus the set of all generalized \mobius transformations, parametrized by 
the set of compatible involutions $\Gamma_{\cal P}$, forms a groupoid.
Now we construct a map between the two Fredholm modules $\alpha_j$, 
which are lifts of the given polarized module $\calp$. 
\bproposition
Put
\beqn\label{F-transform}
g_{21}(F_1)= (\cala F_1 + \calb)(\cala + \calb F_1)^{-1}.
\eeqn
Then $g_{21}(F_1) = F_2$. 
\eproposition
\proof Let us denote by $\cals_{\pm}(F)$ the 
$\pm 1$ eigenspaces of the involution $F$. By definition of $F_j$, we have
that 
$$
\begin{array}{rcl}
\cals_+(F_j) & = & \cale\\
\cals_-(F_j) & = & \gamma_j(\cale)
\end{array}
$$
for $j =1,2$. It follows that $\cals_+(F_1) = \cals_+(F_2)$ and 
$\gamma_2\gamma_1\cals_-(F_1) = \cals_-(F_2)$. If we use expressions for 
$\cala$ and $\calb$, these relations can be expressed as 
$$
\begin{array}{rcl}
(\cala + \calb) \cals_+(F_1) & = & \cals_+(F_1)\\
(\cala - \calb)\cals_-(F_2) & = & \cals_-(F_2).
\end{array}
$$
In turn, these two identities can be combined to give
$$
(\cala +  \calb F_1)\cals_{\pm}(F_1) = \cals_{\pm}(F_2)
$$
which is equivalent to 
$$
F_2 = (\cala + \calb F_1)F_1(\cala + \calb F_1)^{-1}.
$$
This gives the formula \ref{F-transform}, if we use that $F_1$ is an involution. 
\qed

Let $g_{21}$ be the matrix defined in \ref{gen-mobius} whose polar decomposition is given 
by \ref{new-polar}. We define the following action of $g_{21}$ on the Fredholm module 
$\alpha_1$: 
$$
g_{21}(\alpha_1) = (W(\bh_1), W\pi W^{-1}, W\gamma_1 W^{-1}, g(F_1))
$$
where $g(F_1) $ has the same meaning as in \ref{F-transform}. We claim that 
$g_{21}(\alpha_1) = \alpha_2$. 

We  have already checked that $W: \bh_1 \ra \bh_2$ is a unitary operator and that
$g_{21}(F_1) = F_2$. We need to show that   $W\gamma_1 W^{-1} = \gamma_2$ and that 
$W\pi W^{-1} = \pi $

To prove the first statement we use the Cayley transform to write
$Q = (1-m)(1+m)^{-1}$ which together with Proposition \ref{Cayley-transform}
gives
$$
\begin{array}{rcl}
W\gamma_1 W^{-1} & = & \left(\displaystyle\frac{1-m}{1+m}\right)^{\half}
(1+m )\gamma_2 (1+m)^{-1}
\left(\displaystyle\frac{1-m}{1+m}\right)^{-\half}\\
\rule{0mm}{5mm}
& = & (1-m^2)^{\half} \gamma_2 (1-m^2)^{-\half}.
\end{array}
$$
Since $\gamma_2$ anticommutes with $m$, it commutes
with $m^2$, and  by the continuous functional calculus
it also commutes with $(1-m^2)^{\half}$. This gives the required 
result. The second statement follows directly 
from Proposition \ref{Cayley-transform} and the spectral mapping theorem. 

Hence we arrive at the following result.

\bproposition
Let
$\alpha_j = (\bh_j, \pi, \gamma_j, F_j)$, 
$j= 1,2$, be two Fredholm module lifts of a given polarized module 
$\calp = (\calh , \cale, \pi)$. Then there exists a generalized \mobius 
transformation 
$$
g_{21} = \larray \cala & \calb \\
\calb & \cala \rarray = 
\larray \half(1+\gamma_2\gamma_1) & \half(1 - \gamma_2\gamma_1)\\
\rule{0mm}{5mm} \half(1 - \gamma_2\gamma_1 )& \half(1+ \gamma_2\gamma_1)
\rarray
$$
such that $g_{21}(\alpha_1) = \alpha_2$. 
\eproposition
If we now use this Proposition together with the fact that 
generalized \mobius transformations form a groupoid, as described 
in Proposition \ref{groupoid}, we obtain the proof of the 
following description of the action of generalized \mobius transformations
on the space of lifts of the polarized module \calp. 
\btheorem
Let $\calp = (\calh , \cale, \pi)$ be a polarized module. The space of 
all lifts $\alpha= ( \bh, \pi, \gamma, F)$ of \calp\ to a Fredholm module, 
parametrized by compatible involutions $\gamma$ from $\Gamma_{\cal P}$,
forms a category in which morphisms are provided by invertible generalized
\mobius transformations \ref{gen-mobius}.
\etheorem

\pagebreak

\mysubsection{Real structure}

Many examples demonstrate that
Fredholm modules are readily associated
with $C^*$ algebras of complex valued functions on a manifold. 
However, Fredholm modules that can be associated with a real 
smooth manifold are equipped with additional structure which we
shall call the {\em real structure}. 

A real structure on an involutive  algebra $A$ is a conjugate linear map 
$c : A\ra A$ with the properties 
 $c^2 = 1$; 
 $c(ab) = c(a) c(b) $, for all $a, b$ in $A$;
 $c(a^*) =  (c(a))^*$. 

\bdefinition\label{realFred}
An even   Fredholm module 
$\alpha = (\bh, \pi, F, \gamma)$ over an algebra $A$ equipped with a real 
structure is called {\em real } iff it possesses a conjugation operator
$C : \bh \ra \bh$ with the following properties: 
\begin{enumerate}
\item $C$ is conjugate linear and $C^2 = 1$; 
\item $C$ is antiunitary, i.e., $(Cx, Cy) = ( y,x)$ for all $x$, $y$
in $\bh$; 
\item $C\gamma = \epsilon \gamma C$, with $\epsilon $ a fixed factor equal 
$+1$ or $-1$; 
\item $C\pi(a)C = \pi(c(a))$ for all $a$ in $A$. 
\item $CF = FC$. 
\end{enumerate}
\edefinition
Our main example of a real Fredholm module will be associated with 
a conformal manifold of even degree. 

In the same vein we can introduce real structure on a polarized 
module. 
\bdefinition\label{realPol}
 A polarized module
$(({\cal{H}}, \sigma(   \cdot , \cdot )    ), {\pi}, {\cal{E}})$
over an algebra with real structure ${\cal{A}}$
is called {\em real}\,
iff it possesses a conjugation operator
$C : {\cal{H}} \rightarrow {\cal{H}}$
with the following properties:
\begin{enumerate}
\item  $C$ is conjugate-linear and has $C^2 = 1$,
\item  $\sigma(  Cx, Cy)     = {\epsilon}\sigma(  y, x)    $,
for all $x$ and $y$ in ${\cal{H}}$,
with a fixed factor ${\epsilon}$ which is either $+1$ or $-1$.
\item  $C{\pi}(a)C= {\pi}(c(a))$, for all $a$ in ${\cal{A}}$,
\item  $C({\cal{E}}) = {\cal{E}}$.
\end{enumerate}
\edefinition
\bproposition
 Let
$\alpha =({\cal{H}}, {\gamma}, {\pi}, C)$
be a  Fredholm module.
Then the underlying polarized module $\calp(\alpha) $ is also real.
\eproposition
\proof The proof is a simple matter of comparing the definitions. \qed

\pagebreak

\mysection{ Conformal structures and Fredholm modules}
It has been shown by Connes, Sullivan and Teleman  \cite{CST1}\cite{CST2}
that the conformal structure on a compact oriented even dimensional 
manifold correspond to Fredholm modules over the algebra of smooth 
functions on that manifold, and 
in particular that such conformal structures can be reconstructed
from the purely algebraic data of the Fredholm module. The formalism
that has been described in the preceding section can be seen 
as a general approach to noncommutative conformal geometry. 
From this point of view,  
a polarized module $\calp$  encodes certain features of a 
\lq noncommutative differential manifold\rq whereas the set of 
Fredholm modules lying over 
$\calp$ describes conformal structures on this manifold.
The \mobius group, acting on the space of Fredholm modules, 
transforms the corresponding conformal structures in a way 
that cannot be achieved using orientation preserving diffeomorphisms. 
To appreciate the possibilities of this point of view it may be 
useful to see in some detail how the commutative example from ordinary 
differential geometry fits in this general algebraic framework. 

Let $V$ be an oriented compact $2l$-dimensional smooth manifold without 
boundary, and let $\cala$ be the algebra of smooth complex-valued functions over $V$. Let 
us denote by $\calh$ the space of smooth sections of the bundle
$\Lambda^l(V, T^*_C V)$ of complexified $l$-forms on $V$. 
$\calh$ is a locally convex topological space whose completion, 
which we shall also denote $\calh$ is a Hilbertian space. 
Inside $\calh$ there is a canonically determined subspace 
${\rm Im}d$, which is spanned by forms of the type $d\omega$, 
where $d$ is the de Rham differential. Let us denote by 
$\cale$ the closure of this space in $\calh$. The 
space $\calh$ is also equipped with a canonical continuous
nondegenerate hermitian form
$$
\sigma( \omega , \eta ) = (-i)^l \int_V \omega \wedge \bar{\eta}
$$

\blemma
The space $\cale$ is a closed subspace, which is  totally isotropic 
with respect to the indefinite form $\sigma(\cdot, \cdot )$. 
\elemma
\proof We have
$$
\int_Vd\omega \wedge \bar{d \eta} = \int_V d(\omega \wedge \bar{d\eta})
= \int_{\partial V} \omega \wedge \bar{d\eta} = 0.
$$
The result now follows by the continuity of the hermitian form $\sigma(\cdot, 
\cdot )$. 
\qed

The algebra $\cala$ acts on the space of forms by left multiplication, 
as does the $C^*$-algebra $A$ of continous complex-valued functions
on $V$. Let us denote this action by $\pi$. This data gives our most
important example of a polarized module.
\btheorem\label{associated-polmod}
The triple $\calp_{can}=(\calh, \cale , \pi)$ is a polarized module which 
is canonically associated with the manifold $V$. 
\etheorem

We stress here that to construct this polarized module we only used the
differentiable structure of the manifold $V$. It is clearly an interesting 
question to discuss possible lifts of this module to a Fredholm module, 
using the lifting procedure described in the previous section. 
We recall that any such lift was constructed using a compatible 
involution $\gamma$ from the set $\Gamma_{\cal P}$, and so as a first 
step we should identify involutions $\gamma$ that are compatible with 
$\calp_{can}$. 

A Riemannian metric $g$ on $V$ is a smooth, nondegenerate symmetric tensor
of order two which is positive definite everywhere. A {\em 
conformal structure} on $V$ is by definition an equivalence 
class of metrics, where $g_1 \sim g_2$ iff $g_1 = f g_2$ 
for some smooth real-valued function which is everywhere positive.
A choice of a conformal structure $[g]$ on $V$ gives rise to a Hodge 
$*$ operator $* : \calh \ra \calh$ which is used 
to define a scalar product 
$$
(\omega , \eta ) = \int_V \omega \wedge *\bar{\eta}
$$
We denote by $\bh$ the space $\calh$ equipped with this inner product. 
The Hodge operator is an involution up to a 
sign and so we can define the operator $\gamma = (-i)^l *$ which is an involution. Moreover, 
$\gamma$ is self-adjoint with respect to 
both the indefinite form $\sigma( \cdot , \cdot )$ and 
the inner product $(\cdot , \cdot )$. 
\blemma\label{compatible}
The involution $\gamma$ is a compatible involution associated 
with the polarized module $\calp_{can}$. 
\elemma
\proof This follows directly from the known properties of the 
Hodge $*$-operator. \qed

The involution $\gamma$ gives the familiar orthogonal decomposition
of $\bh$ into spaces of self-dual and anti-self-dual forms: 
$$
\bh = \bh_0 \oplus \bh_1. 
$$
Let us now assume that the manifold $V$ has trivial cohomology in 
the middle dimension, which means that there are no harmonic $l$-forms on 
$V$. Then there is another decomposition of the space $\bh$, namely
$$
\bh = \cale \oplus \gamma(\cale)
$$
Define $F$  to be the operator which $+1$ on $\cale$ and $-1$
on $\gamma(\cale)$. The following result is then a direct consequence of 
our lifting procedure. 

\btheorem
 The operator $F$ is a self-adjoint involution anticommuting with $\gamma$. 
The data $\alpha = (\bh, \gamma, \pi, F)$ determine a Fredholm module associated
with the conformal structure $[g]$, which is a lift of the canonical polarized module
$\calp_{can}$.
\etheorem

We remark that the Fredholm module constructed here is a special case of the
Fredholm module defined by Connes, Sullivan and 
Teleman for a general even-dimensional  conformal manifold \cite{Connes:Book,CST1,CST2}.

As this theorem indicates, Fredholm modules contain information about 
the conformal structure of the manifold. 
The \mobius group, acting on the the space of Fredholm modules associated
with a differentiable manifold through the various choices of conformal 
structure, deforms the underlying conformal structures in a way that 
is impossible to obtain using orientation preserving diffeomorphisms. 
Thus it seems plausible that \mobius transformations would have 
some relation to the deformation theory of conformal structures (see, e.g.
\cite{KK}). This point will need to be investigated further. 

In the case when $V$ has nontrivial cohomology in the middle dimension, 
the above construction produces a {\em pre-Fredholm module} in the 
sense that the operator $F$ is not an involution but has the 
property that $1-F^2$ is a compact operator. (In fact it is of  finite rank 
in this case.) As was shown by Connes in his seminal paper \cite{Connes:BigPaper}, 
any pre-Fredholm module can be lifted to a Fredholm 
module, thus yielding a Fredholm module associated with a conformal 
structure on $V$. Fredholm modules are clearly useful especially when 
one wants to construct the corresponding characters (cyclic cocycles), 
but it transpires  that  pre-Fredholm modules  already contain
interesting geometric information. There is a modification of the 
theory of \mobius transformations that fits that case, and it 
will be developed in a forthcoming paper.

\bremark 
The polarized and Fredholm modules constructed in this section are 
in fact {\em real}, in the sense of Definitions \ref{realFred}
and \ref{realPol}.
\eremark
\bremark
To construct a conformal structure from a given Fredholm module, 
Connes \cite[p.~334]{Connes:Book} uses the following formula to calculate
$L^{2l}$-forms of one-forms on $V$: 
$$
\Tr_\omega \left( \left| \sum f_i [ F, g_i]\right|\right)
= c_l \int_V \norm{\sum f_i dg_i}^{2l}
$$
Here $\Tr_\omega$ is the Dixmier trace.
We want to show that unitarily equivalent Fredholm modules give the 
same conformal class. But this is clear from our localization 
theorem \ref{local}, which shows that it is sufficient to prove this statement
for unitary operators that belong to the commutant of a fixed representation of 
the algebra $\cala$. In this case the unitary $m\in \calu(M)$ acts by 
$$
\begin{array}{rcl}
\Tr_\omega \left(\sum f_i[F, g_i]\right) & \mapsto&  \Tr_\omega \left(
\sum f_i [mFm^{-1},  g_i] \right) \\
& = &  \Tr_\omega \left(m\sum f_i[F, g_i] m^{-1} \right) \\
& = & \Tr_\omega \left(\sum f_i[F, g_i]\right)
\end{array}
$$
Thus the norm, and therefore the associated conformal class, 
does not change under this transformation. 
\eremark

\end{document}